\documentclass[aps,preprint,amsmath,showpacs,superscriptaddress]{revtex4}
\usepackage{epsfig}
\begin{document}
\renewcommand{\baselinestretch}{1.0}
%\preprint{}

\title{Spin states and phase separation in La$_{1-x}$Sr$_{x}$CoO$_3$
($x=0.15, 0.25, 0.35$) films: optical, magneto-optical and magneto-transport
studies}



\author{N. N. Loshkareva}
\affiliation{Institute of Metal Physics of Ural Division of RAS,
Ekaterinburg, 620219, Russia}

\author{E. A. Gan'shina}
\affiliation{Moscow State University, Moscow, 119899, Russia}

\author{B. I. Belevtsev}
\email[]{belevtsev@ilt.kharkov.ua}
%\thanks{}
\affiliation{B. Verkin Institute for Low Temperature Physics and Engineering,
National Academy of Sciences, Kharkov 61103, Ukraine}

\author{Yu. P. Sukhorukov}
\affiliation{Institute of Metal Physics of Ural Division of RAS,
Ekaterinburg, 620219, Russia}

\author{E. V. Mostovshchikova}
\affiliation{Institute of Metal Physics of Ural Division of RAS,
Ekaterinburg, 620219, Russia}

\author{A. N. Vinogradov}
\affiliation{Moscow State University, Moscow, 119899, Russia}

\author{V. B. Krasovitsky}
\affiliation{B. Verkin Institute for Low Temperature Physics and Engineering,
National Academy of Sciences, Kharkov 61103, Ukraine}

\author{I. N. Chukanova}
\affiliation{Institute for Single Crystals, National Academy of Sciences,
Kharkov 61001, Ukraine}


%\date{\today}

\begin{abstract}
Optical absorption and transverse Kerr effect spectra,   resistivity and
magnetoresistance of  La$_{1-x}$Sr$_{x}$CoO$_3$  ($x=0.15, 0.25, 0.35$)
films have been studied. The temperature dependencies
of the optical and magneto-optical properties of the films exhibit features,
which can be attributed to the transition of the Co$^{3+}$ ions from the
low-spin state ($S=0$) to the intermediate-spin state ($S=1$) and to orbital
ordering of the Co$^{3+}$ ions in the latter state. The evolution of the
properties influenced by doping with Sr is interpreted on the basis of the
phase separation model.
\end{abstract}


\maketitle
\normalsize{
\section{Introduction}
Discovery of the so-called colossal magnetoresistance in manganite films
\cite{helm} has renewed an interest in other ferromagnetic perovskite-like
compounds. These include LaCoO$_{3}$-based cobaltites which are in many ways
similar to manganites and yet they have some fundamental distinctions. At
low temperatures, LaCoO$_{3}$ is a non-magnetic insulator. In this state
Co$^{3+}$ ions are mainly in the low-spin (LS) state ($t_{2\mathrm{g}}^{6},
S=0$) \cite{raccah} because the crystal-field energy  dominates slightly
over the intraatomic Hund's energy \cite{raccah,korotin}. However, as the
temperature rises, Co$^{3+}$ ion state changes gradually from the LS state
to the high-spin (HS) state ($t_{2\mathrm{g}}^{4}e_{\mathrm{g}}^{2}, S=2$)
or to the intermediate-spin (IS) state
($t_{2\mathrm{g}}^{5}e_{\mathrm{g}}^{1}, S=1$). The latter scenario is
supported by the theoretical calculations \cite{korotin,ravin} and results
of some experimental studies \cite{saito, yamaguchi3,kobayashi,zobel}. The problem
of spin states of Co$^{3+}$ ions in cobaltites is not settled to date and
remains topical.
\par
When LaCoO$_{3}$ is doped with Sr$^{2+}$ ions, hole-rich regions appear in
the hole-poor matrix \cite{senaris1}. Evolution of the magnetic properties
as a function of Sr$^{2+}$ concentration $x$ and temperature is governed by
the transition from the spin-glass-like to cluster-glass state or to the
ferromagnetic (FM) metallic state \cite{caciuf,itoh}. The certain scatter in
the estimates obtained by different authors \cite{senaris1,caciuf,itoh} for
the critical Sr$^{2+}$
concentration at which long-range FM order sets in is accounted for by
such factors as instability against the formation of clusters rich in
Sr$^{2+}$ and containing ions in the IS state, the dependence of the chemical
heterogeneity upon the thermal history of the samples and the degree of
oxidation. The question is still discussed considerably whether clusters
are  generated by compositional inhomogeneities when, for example, the
phases with $x=0.5$ and $x=0.2$ are present  in
La$_{0.85}$Sr$_{0.15}$CoO$_{3}$ \cite{anil} or they appear due to electron
phase separation \cite{mira}.
\par
In this article, a study of the Kerr effect, optical
absorption and transport phenomena in La$_{1-x}$Sr$_{x}$CoO$_3$
($x=0.15, 0.25, 0.35$) films is presented. The phase separation in films
can be expected to be somewhat more complicated as compared with bulk
materials since it could depend on strain induced by the film-substrate
lattice mismatch. At the same time, films are suitable objects
which can be investigated by traditional electric and magnetic techniques
giving averaged characteristics of the (inhomogeneous) material and by
optical methods permitting to separate responses from conducting regions and
the insulating matrix and to study the magnetic processes inside the
conducting clusters. This combined approach was applied earlier to
manganites \cite{loshka1,loshka2}: the charge inhomogeneities were studied
by the optical absorption technique while the magnetically inhomogeneous
state was  investigated using magneto-optical (MO) method.
\par
Transport and  optical properties of FM oxides are known to depend
significantly on their magnetic properties. As for the MO properties,
they are determined directly by the magnetization of a material studied.
In this work, the MO properties were measured using the transverse Kerr
effect (TKE).
Generally, the MO Kerr effect consists in an influence of the magnetization
of the medium on the reflected light. The TKE consists in the intensity
variation of a light reflected by a magnetized sample in conditions where
magnetic field is applied parallel to the sample surface and perpendicular
to the light incidence plane.
Measurements of MO spectra in the visible and near ultra-violet
spectral range, i.e. in the range of the fundamental absorption,
could provide an information concerning electronic structure of the
cobaltites and its dependence on chemical composition,
conditions of the synthesis and other factors. In particular, the
appearance of the new type of magnetic ions, Co$^{4+}$, due to doping with
Sr, and/or change in the spin state of Co ions are expected to influence the
MO spectra. The magnitude of the MO response is known to be the product
of spin-orbit coupling strength and net electron spin polarization
(magnetization). This makes the magneto-optical
effects sensitive to the magnetic state of unfilled $d$-shells in the
transition-metal ions. MO spectroscopy provides not only the information
about the total density of states (similarly to normal optical
measurements) but also about the electron spin polarization of states
participating in the magneto-optical transition \cite{schoenes,zvezdin}.
\par
The temperature dependence for different MO effects measured at a definite
wavelength indicates the variation of the magnetic order in the samples.
Temperature and magnetic field dependences of MO effects reveal the phase
transition temperatures and peculiarities of the originated magnetically
ordered states. MO response is sensitive not only to the long-range magnetic
order but even to the short-range magnetic order. Thus, formation of the
ferromagnetic clusters would manifest itself in the MO properties.
\par
The phenomenological description of  MO effects is based on consideration of
the influence of a magnetic field on the dielectric permittivity tensor,
$\varepsilon_{ij}$, of the medium \cite{zvezdin}. If the tensor is
symmetric ($\varepsilon_{ij}=\varepsilon_{ji}$) in zero magnetic field
$H=0$, it becomes non-symmetric [$\varepsilon_{ij}(H)=\varepsilon_{ji}(-H)$]
in a nonzero magnetic field. In the linear approximation, the dielectric
permittivity tensor of the gyroelectric medium is
\begin{equation}
\tilde{\varepsilon}=
\left\{
\begin{matrix}
\varepsilon_{0} & \mathrm{i}\varepsilon_{xy} &  0\\
-\mathrm{i}\varepsilon_{xy} & \varepsilon_{0} &  0\\
0  & 0 &  \varepsilon_{0}
\end{matrix}
\right\},
\label{eq1}
\end{equation}
where the diagonal elements,
$\varepsilon_{0} = \varepsilon_{0}^{'}-\mathrm{i}\varepsilon_{0}^{''}$,
describe normal optical properties (which do not depend on magnetization)
and off-diagonal elements,
$\varepsilon_{xy} = \varepsilon_{xy}^{'}-\mathrm{i}\varepsilon_{xy}^{''}$,
proportional to magnetization, are related to MO properties.
All MO effects, which are linear in magnetization, can be expressed in terms
of  off-diagonal elements of the dielectric permittivity tensor.
\par
For the TKE, variation of the reflected light intensity for $p$-polarized
wave due to the magnetization of a ferromagnetic sample can be written as
\cite{zvezdin}
\begin{equation}
\delta_{\mathrm{p}} =
2\sin(2\varphi)\frac{A_{1}}{A_{1}^{2}+B_{1}^{2}}\,\varepsilon^{'}_{xy} +
2\sin(2\varphi)\frac{B_{1}}{A_{1}^{2}+B_{1}^{2}}\,\varepsilon^{''}_{xy},
\label{eq2}
\end{equation}
where $A_{1}=\varepsilon_{0}^{''}[2\varepsilon^{'}_{0}\cos^{2}(\varphi)-1]$,
$B_{1}=[\varepsilon_{0}^{''2}-\varepsilon_{0}^{'2}]\cos^{2}(\varphi)+
\varepsilon^{'}_{0} - \sin^{2}(\varphi)$, $\varphi$ is incidence angle
of light.
\par
The results presented in this article can be taken as supporting the
transition of the Co$^{3+}$ spins from the LS to the IS state with
increasing temperature. Beside this, the results agree with the presence
of the orbital ordering of the Co$^{3+}$ ions in the IS state.
At consideration of changes in the properties of the films with
Sr doping the phase-separation effect is taken into account.

\section{Samples and experimental techniques}
The films (about 200 nm thick) were grown by pulsed-laser
deposition onto a (001)-oriented LaAlO$_3$ substrate with a 1.06 $\mu$m
Nd - YAG laser (pulse length 10 ns, pulse energy 0.33 J, pulse frequency
12 Hz). The ceramic targets  with nominal compositions
La$_{1-x}$Sr$_{x}$CoO$_3$ ($x=0.15, 0.25, 0.35$) were prepared by
solid-state reaction method. We were aware that available  methods of
chemical control can determine actual compositions in targets and films
with an accuracy not better than $\pm 2$ atomic percent. Besides, the
composition of pulsed-laser deposited films can differ slightly from that of
the target. For all these reasons, the nominal compositions were chosen in
such way that one of them ($x=0.15$) is
sure below the percolation threshold ($x_{\mathrm{c}}=$~0.20--0.25)
\cite{senaris1,caciuf,itoh}, at which the infinite FM cluster is formed below
the Curie temperature in this system, the second ($x=0.25$) should be
near the percolation threshold, and the third ($x=0.35$) is sure above
this threshold. The results outlined in the following sections show that we
have achieved this aim to a great extent.
\par
At the ablation of target material, the
substrate was shielded to avoid its direct exposure to the plasma plume.
In this case the deposition occurs from the flux that is reflected from a
side screen. This has ensured a high surface smoothness, which is essential
for obtaining reliable results at an optical study.  The deposition was
performed in oxygen atmosphere at the pressure 8 Pa and the substrate
temperature 880$\pm$50$^{\circ}$C. The deposited films were cooled to room
temperature in oxygen at 10$^5$ Pa (1 atm). These preparation
conditions make possible coherent epitaxy. The investigations of other
cobaltite films, prepared in the same pulsed-laser deposition chamber and
in the same conditions, provide reason enough to suggest that the film
studied are polycrystalline, even if rather highly oriented.
\par
The optical absorption spectra of the films were measured at 80-295 K with a
prism infrared spectrometer (in the range 0.1--1.4~eV) and a grating
spectrometer (in the range 1.0--4.0 eV).
The measurements of TKE were made using an automatic MO spectrometer
\cite{balyk}. A dynamic method to record TKE was used. The relative
change in the intensity of the reflected light $\delta = [I(H)-I(0)]/I(0)$,
where $I(H)$ and $I(0)$ are the intensities of the reflected light in the
presence and in absence of a magnetic field respectively, was directly
measured in the experiment. The magnitude of the alternating magnetic field
in the gap of an electromagnet was up to 3.5 kOe. The relative precision
of the apparatus is $10^{-5}$. MO spectra were recorded in the photon
energy range 1.3--3.8 eV at a fixed light incidence angle
$\varphi = 67^{\circ}$. For MO measurements, a continuous-flow helium
cryostat is used, which makes possible investigations in the temperature
range 10-300 K.
\par
The resistivity of the films was measured as a function of temperature and
magnetic field (up to 20 kOe) using the standard four-point probe
technique. The field was applied parallel or perpendicular to the film
plane. In both cases it was perpendicular to the transport current.

\section{Experimental results}
\subsection{Kerr effect}
The temperature dependences of TKE measured on
La$_{1-x}$Sr$_{x}$CoO$_3$ films (in the magnetic field $H=3.5$~ kOe and at
photon energy $E=2.8$~eV) are shown in Fig. 1. The $\delta(T)$ curves reveal
characteristic features at the transition to the magnetically ordered state.
The position of the abrupt decrease in TKE on the temperature scale (when
going from lower to high temperature) corresponds to the Curie temperature
$T_{c}$. As follows from given curves, Curie temperatures for films with
$x=0.35$ and $x=0.25$  coincide and are equal to 230 K. Below 230 K the
films with $x\geq 0.25$ exhibit a considerable magneto-optical effect with
a peak at $T\approx 180$~K. As temperature decreases further, TKE drops
sharply: its value becomes more than three times smaller in the range
110--160 K for $x=0.35$ or in the range 130--175~K for $x=0.25$. For the
film with $x=0.15$, TKE has a peak at $T=118$~K and another anomaly
at $T\approx 60$~K. Note that in the low temperature region ($T<118$~K),
TKE for $x=0.15$ is larger than for films with higher Sr concentrations.
The $\delta (T)$ shape is dependent considerably on measurement conditions.
As seen in Fig. 1, the TKE curves for $x=0.15$ and $x=0.35$ measured on
cooling and heating in a high field ($H=3.5$~kOe) are shifted relative to
each other. Hysteresis phenomena were much smaller for the film with
$x=0.25$ and were observed only near the maximum.
\par
The position of the maximum in the TKE temperature  curve is dependent on
the applied magnetic field and shifted to the high-temperature region when
the field decreased. On heating the film with $x=0.35$ in the field
$H=0.9$~kOe, the $\delta (T)$ curve shifts drastically (over 70 K) towards
higher temperatures [Fig. 1(c)] when compared with the $\delta (T)$ curve
taken in the higher field $H=3.5$~kOe. Both the curves nearly coincide at
$T>200$~K. On a fast cooling (with a rate about 20 K/min) of the film with
$x=0.25$, the $\delta (T)$
curve shifts to the high-temperature region enormously and the maximum
magnitude of TKE decreases greatly [Fig. 1(b)]. At $T<110$~K, TKE for films
with $x\geq 0.25$ decreases linearly with temperature (Fig.~1).
\par
The spectral dependences of TKE measured at temperatures of the maximum in
their temperature curves are shown in Fig.~2. It is seen that the spectra
are similar for all Sr concentrations studied. The Kerr effect has a positive
maximum at $E\approx 1.5$~eV and a broad negative maximum in the range
of $E\approx 2.7$~eV. The amplitudes of the peaks increase with the growing
Sr concentration. Above 2.0 eV the spectra exhibit a fine structure which
is most distinct for the film with $x=0.15$ (peculiar features at 2.1, 2.6
and 3.1 eV stand out against the background of the broad maximum).
\par
The field dependences of TKE, $\delta(H)$, for low-temperature range are
linear for all films studied (Fig.~3). They are, however, quite different
near the temperatures corresponding to maximum in $\delta(T)$ dependences.
For the film with $x=0.15$ the
$\delta (H)$ is linear up to 3.5 kOe, whereas for films with $x=0.25$ and
$x=0.35$ it looks like a field-magnetization curve for FM compounds.

\subsection{Resistivity and magnetoresistance}
\label{resist}
The temperature dependences of resistivity, $\rho (T)$, for the films
studied are presented in Fig. 4 in semilogarithmic coordinates.
Since resistivity of the film with $x=0.35$ shows only small visible
temperature variation in this scale, it is presented more clearly in Fig.~5
in linear coordinates. Dependences $\rho (T)$ for films with $x=0.15$ and
$x=0.25$ have non-metallic character ($\mathrm{d}\rho/\mathrm{d}T<0$)
(Fig.~4). They are found to be rather unusual: $\rho (T)$ follows closely
the dependence $\rho (T)= \rho (0)\exp(-aT)$ for $x=0.15$ and is practically
linear in temperature for $x=0.25$. The dependence $\rho (T)$ for $x=0.35$
has a maximum at $T\approx 250$~K and a shallow minimum at $T\approx 40$~K
(Fig.~5). The minimum is most probably connected with a polycrystalline
structure of the film \cite{belev}. The behavior of $\rho (T)$ is metallic
between these extremum points.
\par
The temperature dependences of magnetoresistance (MR) $\Delta R(H)/R(0)=
[R(H)-R(0)]/R(0)$ measured in the magnetic field $H=20$~kOe are shown in
Fig.~6. It is seen that for this field the MR is negative for all films
studied. The dependence taken on the sample with $x=0.35$ is quite typical
of optimally doped FM cobaltites and manganites [Fig.~6(a)]:
there is a quite pronounced peak (in the MR modulus) near the Curie
temperature $T_{\mathrm{c}}\approx 230$~K; it goes down for temperature
deviating to either side from $T_{\mathrm{c}}$. This type of MR behavior is
characteristics of so called intrinsic mechanism of MR, which depends
on the magnetic order (an external field enhances the magnetic
order that, in its turn, leads to a decrease in the resistivity and,
therefore, to negative MR). This mechanism works in the FM state only. For
this reason the MR decreases to zero for temperatures above $T_{\mathrm{c}}$.
Far below $T_{\mathrm{c}}$, the MR of this type goes to zero as well
since an applied magnetic field can not strengthen the magnetic order
appreciably in low temperature range ($T\ll T_{\mathrm{c}}$) where the
magnetization is close to saturation.
\par
In a magnetically inhomogeneous sample,
consisting of weakly connected FM regions in dielectric or non-magnetic
matrix, the so called extrinsic mechanisms of MR can give an additional
contribution to the total MR. These mechanisms are determined by
the charge transfer between the poorly connected or isolated FM regions.
This type of magnetic inhomogeneity in FM
oxides can be connected with grain boundaries in polycrystalline
samples or with the phase separation effects. A contribution of the extrinsic
MR increases with temperature decreasing and is maximum at zero temperature.
A concurrence of the intrinsic and extrinsic MRs can result in non-monotonic
temperature dependence of the total measured MR in cobaltite films
\cite{belev,belev2}.
Discussion of possible mechanisms for the extrinsic type of MR in
inhomogeneous polycrystalline samples can be found in Refs. \cite{ziese}. It
is clear that these mechanisms are applicable (at least generally) also in
the case of the phase separation when this results in a system of hole-rich
FM clusters embedded in hole-poor dielectric matrix.
In this way, the temperature behavior of MR in FM oxides can reflect their
magnetic homogeneity. Taking it all into account,
it can be said from Fig. 6(a) that film with $x=0.35$ is the most
homogeneous from the films studied although even it has appreciable
MR at low temperature range that indicate the presence of some magnetic
inhomogeneity induced by one or both of the above-mentioned reasons.
The MR of the film with $x=0.25$ increases continuously with temperature
decreasing [Fig. 6(b)] what corresponds to behavior of highly
inhomogenenous system. The same is true in relation to the films with
$x=0.15$ where additionally a non-monotonic temperature behavior of MR
can be seen. It should be noted
that the largest MR is observed in the least
doped sample ($x=0.15$) [Fig.~6(c)]. For the films with $x=0.15$
and 0.25, the temperature interval, where MR increases with lowering
temperature, is fairly close to the temperature range of increasing TKE.
\par
The MR of the films with $x=0.35$ and $x=0.15$ is anisotropic (Fig.~6).
The absolute values of negative MR are much higher
in fields parallel to the film plane than in perpendicular ones.
The anisotropic MR is not surprising for pulsed-laser deposited films of
FM perovskite oxides. For cobaltites, this effect is seen earlier
in La$_{0.5}$Sr$_{0.5}$CoO$_3$ film \cite{belev}. Since the conductivity
in mixed-valence cobaltites increases with an enhancement of the FM
order, this behavior just reflects the point that the magnetization
increases more easily in a magnetic field parallel to the film plane. It
is connected mainly with the shape anisotropy of the magnetization.
The strain-induced anisotropy due to the film-substrate lattice interaction
can have an influence as well. The influence of
this type of MR anisotropy  has been found
in La$_{0.5}$Sr$_{0.5}$CoO$_3$ film \cite{belev2}. All these types of the
MR anisotropy are associated with FM state and, for this reason,
disappear  above $T_{\mathrm{c}}$. In the films studied,
MR anisotropy is found only below $T\approx 200$~K for $x=0.15$ and below
$T\approx 250$~K for $x=0.35$. The highest MR anisotropy is observed
in the film with $x=0.15$ [(Fig.~6(c)]. No appreciable MR anisotropy is
found in the film with $x=0.25$.
\par
Generally, the MR curves of the films studied have been hysteretic with
specific structures in low fields (see Fig.~7 for $x=0.35$ at $T=78$~K).
It is well established that the behavior of the MR curves for FM oxides
correlates with that of magnetization curves \cite{ziese}. In particular,
the field $H=H_{\mathrm{p}}$, where the resistance reaches its maximum
(Fig.~7), corresponds to the coercive force $H_{\mathrm{c}}$.
It is found that field $H_{\mathrm{p}}$ is maximal at lowest temperature
in this study ($T\approx 4.2$~K) but it drops to zero above
$T\leq 77$~K for $x=0.15$
and above $T\approx 90$~K for $x=0.35$. Above certain field values,
hysteresis no longer occurs, which implies that magnetization reversal
processes start in these fields. Such fields are much lower for
the parallel orientation (Fig.~7). We will discuss the hysteresis
phenomena more elaborately below.

\subsection{Optical absorption}
The optical density spectra $D=\ln(I_{0}/I) \equiv \ln(1/t)$ ($I_0$ is the
incident-light intensity, $I$ is the intensity of light transmitted through
the films, $t$ is the transmittance) in the visible and infrared (IR) ranges
are shown in Figs.~8 and 9, respectively. The optical density spectrum
(corresponds to absorption spectrum where the reflection and film thickness
was not taken into account)
for the film with $x=0.15$ has a wide
band at $E\approx 3.0$~eV and quite narrow bands at about 1.15, 1.9, 2.4 eV
in the fundamental absorption region (Fig.~8). When the Sr concentration is
increased, the low-energy band ($1.15\pm 0.03$)~eV shifts slightly towards
higher energies. For $x=0.35$ it is centered at 1.3 eV. The 1.9 eV band
broadens and shifts towards lower energies.
\par
For the film with $x=0.15$, absorption in the IR range goes up for
energy increasing above 0.9 eV (Fig.~9), which indicates the onset of
fundamental absorption.  At
$E<0.9$~eV the light interaction
with charge carriers contributes to the absorption of the films
with $x=0.25$ and $x=0.35$. As follows from the reflection spectra for
LaCoO$_3$ \cite{yamaguchi1}, phonon absorption begins below 0.08 eV.
\par
The absorption in the film with $x=0.15$ is considerably lower at 80 K than
at 295~K in the IR range (Fig.~9). For the film with $x=0.25$
the absorption in the region $E<0.9$~eV is slightly higher at 80 K than at
295 K. For the film
with $x=0.35$ the absorption at $E<0.9$~eV is considerably higher at 80 K
than at room temperature, which implies that the contribution of charge
carriers to the spectrum is larger for $x=0.35$ than for $x=0.25$.
\par
As shown for manganites \cite{loshka1,loshka2,loshka3}, in the energy range
of light interaction with charge carriers, the behavior of $t(T)$ [or
transmitted light intensity $I(T)$]  follows the temperature behavior of
resistivity in the case that metallic FM region
percolates  through the whole material. For the manganites, which for
some reasons consist of separated metallic regions embedded in
semiconducting matrix, the $I(T)$ dependence reflects the temperature
behavior of the resistivity in these isolated metallic regions (or clusters)
and this behavior can be quite different from the direct-current $\rho (T)$
dependences.
The $I(T)$ behavior in Sr-doped cobaltite films at $E=0.20$~eV is
shown in Fig.~10(a). On cooling the transmitted light intensity
(transmittance) of the film with $x=0.15$ increases, i.e. the transmittance
behavior is semiconductive at  80--300 K. In the film with $x=0.25$ the
transmittance exhibits metallic temperature behavior at $T<180$~K, although
no this type of behavior can be seen in the $\rho (T)$ dependence (Fig.~4).
In the film with $x=0.35$ transmittance drops significantly below
$T\simeq 250$~K (metallic behavior), which is in agreement with the
$\rho (T)$ behavior for this film (Fig.~5).
\par
The $I(T)$ dependences of cobaltite films studied have an extra anomaly
(most pronounced for $x=0.35$) in the same temperature
range 160--220 K (Fig.~10). This type of anomaly was never seen in
manganites. The anomaly is observed not only at the photon energy range
where light interaction with charge carriers occurs [Fig.~10(a)], but
at the range of the fundamental absorption edge at $E=$~1.0--1.4~eV
[Fig. 10(b)] as well, where only small ($x=0.25$, $x=0.35$) or  no ($x=0.15$)
contribution of charge carrier to absorption is expected.
The anomaly is more complicated at $E=$~1.0--1.4~eV [Fig. 10(b)] than at
$E=0.20$~eV [Fig. 10(a)]. This may be connected with the intricate character
of the spectrum near the fundamental absorption edge, where the closely
spaced and spectrally unresolved absorption bands overlap. Note that in
our experiments the spectral bandpass of slits in the two spectrometers
used is different in the region of overlapping working ranges. The
spectra in Fig.~8 measured with a smaller spectral bandpass are therefore
resolved better than the spectra in Fig.~9.
\par
The magneto-transmittance effect (analogous to magnetoresistance) detected
in manganite films \cite{sukh} is not found in cobaltite films studied up to
$H=10$~kOe apparently because cobaltites have much lower magnetoresistance.
The influence of magnetic fields on transmittance of the films shows up
itself, however, in rather different behavior of the $I(T)$ curves taken at
$E=0.20$~eV on cooling in zero and finite ($H=8$~kOe) fields. The strong
effect of the sample
thermomagnetic prehistory is illustrated in Fig.~10(c). Curves 1-4 describe
the $I(T)$ dependences under successive changes in measurement conditions.
The $I(T)$ anomaly weakened on field cooling (FC) and measurement in $H=0$
(curve 2) or on zero-field cooling (ZFC) and measurement in $H>0$ (curve 3).
The anomaly is suppressed completely when the film is heated above room
temperature and then cooled quickly (curve 4). Additional explanatory
comments to the measurement conditions can be found in caption to Fig. 10.

\section{Discussion}

\subsection{Absorption spectra and Kerr effect}

There are only scanty optical studies on La$_{1-x}$Sr$_{x}$CoO$_3$
\cite{yamaguchi2,arima,tokura,yamaguchi1,yamaguchi3}. It can be found among
them optical conductivity spectra obtained by the Kramers-Kronig analysis
of reflectivity spectra of poly- and single crystals \cite{tokura}. The
polar Kerr effect
for La$_{1-x}$Sr$_{x}$CoO$_3$ single crystals at room temperature was
studied in Ref. \onlinecite{yamaguchi2}. The optical density spectra $D(E)$
of the films studied have more features in the visible range (Fig. 8) than
the light conductivity spectra $\sigma (E)$ of La$_{1-x}$Sr$_{x}$CoO$_3$
single crystals with similar composition \cite{tokura}. The $\sigma (E)$
spectra of single crystals have three bands at energies about 1, 3,
and 6 eV, respectively \cite{tokura}. They are attributed \cite{solovyev}
to excitations into the $e_g$ band from the $t_{2g}$ band and broad O$(2p)$
band  split by the hybridization effects into the bonding and antibonding
parts. The spectra of the cobaltite films studied have four bands in
energy range 1.0--3.5 eV (Fig.~8). This can be presumably attributed to
superposition of geometric resonances (surface plasmons) induced by
structural inhomogeneities on charge transfer transitions (as shown for
manganites in  Ref. \onlinecite{moskvin}).
Inhomogeneous structures may appear due to the phase separation,
polycrystallinity of the films and strains at the film-substrate boundary.
The evolution of absorption spectra with doping for the films studied
is similar to that of single crystals \cite{tokura}: as the Sr concentration
increases, the spectral weight transfers from the visible range to the IR
region, where light interaction with free charge carriers occurs.
\par
The TKE magnitude is comparable in cobaltites (Fig.~2) and manganites
\cite{yamaguchi2}. The high value of the Kerr effect may be connected
with the strong spin-orbit interaction expected for Co$^{3+}$ (IS) ions.
As for La$_{1-x}$Sr$_{x}$CoO$_3$ single crystal \cite{yamaguchi2}, TKE in
the films studied was
maximum in the broad band near 3.0 eV. In Ref. \onlinecite{yamaguchi2} this
behavior was attributed to the $d$-$d$ transition
$t_{2g}\rightarrow e_{g}^{*}$
allowed for the majority-spin (or spin-up) states because of the
hybridization with O$(2p)$ orbitals. Large  MO effect found in the films
studied may also be related with this transition, which must take place
only for Co$^{3+}$ ions in the IS or HS states, but not in the LS state.
Qualitatively, the TKE behavior can also
be explained on the basis of band calculation for LaCoO$_3$ with Co$^{3+}$
(IS) ions \cite{korotin}. According to the electron-energy scheme for
Co ions in the IS state \cite{korotin}, the magneto-optical effect should
first appear in the spin-up subband and reach its maximum at
$E\approx 2.0$~eV. Then, at $E>3.5$~eV the transitions in the spin-down
subband can be expected. The competition of these transitions can also
account for the change in sign and spectral behavior of the effect.

\subsection{Temperature dependences of transmittance, Kerr effect,
resistivity and magnetoresistance}

It is found for lanthanum  manganites of varying compositions
\cite{loshka1,loshka2} that the temperature at which the FM contribution
appears (estimated from the temperature dependences of TKE) agrees well with
the temperature of the insulator-metal transition in the hole-rich
separated regions or clusters (found from IR absorption data).
A direct correlation of this type is not seen in cobaltite films
studied [compare  Figs.~1 and 10(a)]. We will try in the following
to clear up a possible reason for this discrepancy. In doing so we
can assume that the holes appearing with Sr doping of cobaltites cause phase
separation into the hole-poor matrix and hole-rich regions (e.g., see
\cite{senaris1,caciuf,mira}).

\subsubsection{$\mathrm{La}_{0.85}\mathrm{Sr}_{0.15}\mathrm{CoO}_{3}$ film}
\label{0.15}
Our magneto-optical, optical,  MR and MR anisotropy results for the film with
$x=0.15$ support the outlined general picture for magnetic processes in
cobaltites with low doping level below the percolating threshold
($x_{\mathrm{c}}=$~0.20--0.25) \cite{itoh,senaris1,caciuf}.
The monotonic (except for the range 160--220 K) temperature dependence of
intensity of the transmitted light at $E=0.20$~eV in the film with x=0.15
[Fig.~10(a)] suggests that the clusters with Co$^{4+}$ ions are rather small
in size and their volume fraction is much less than that of the
semiconducting  matrix. For this reason, apparently, the transition of these
isolated clusters to more conductive FM state with temperature decreasing
does not show itself either in  $I(T)$ [Fig. 10(a)] or in
$\rho (T)$ (Fig.~4) dependences.  The TKE value characterizing the FM
contribution is, however, quite high -- only three times less than that
for $x=0.35$. TKE has a maximum at 118 K. After cooling in zero field,
another feature (a shoulder) appears at $T\simeq 60$~K [Fig.~1(a)]. A maximum
at nearly the same temperature $T_{\mathrm{g}}=$~60--70~K was found earlier
in temperature dependences of the AC and DC susceptibility for bulk
La$_{0.85}$Sr$_{0.15}$CoO$_3$ \cite{senaris1,caciuf,anil,mira} and
attributed to  spin-glass freezing. Magnetization temperature curves
for La$_{0.85}$Sr$_{0.15}$CoO$_3$ show a cusp at the same freezing
temperature $T_{\mathrm{g}}$ \cite{itoh,golovanov}. It can be thought that
the above-mentioned feature in the TKE temperature curve near the
temperature $T\approx 60$~K [Fig.~1(a)] is determined by the same effect.
\par
It should be noted that the highest MR and its anisotropy for this film are
observed in the range 60--80~K [Fig.~6(c)]. The ratio of MRs measured in
fields parallel and perpendicular to the film plane is close to 2. At
$T=4.2$~K and 20.4 K the values of field $H_{\mathrm{p}}$  are found to be
rather high ($H_{\mathrm{p}\parallel} = 7.0$~kOe and
$H_{\mathrm{p}\perp}=4.5$~kOe for field orientation parallel and
perpendicular to the film plane, respectively). At $T=78$~K and above the MR
hysteresis is practically unobservable. Taking into account that field
$H_{\mathrm{p}}$ is equal to the coercive force $H_{\mathrm{c}}$ (see
Sec.~\ref{resist} above) this behavior is quite consistent with the assumed
cluster-glass state at this doping level. A cluster glass is actually a
system of small FM (single domain) particles. An isolated particle should
have enough thermal energy to surmount the energy barrier $\Delta E = KV$
($K$ is the anisotropy constant, $V$ is the particle volume) to reverse
its magnetization in magnetic field \cite{chant,kat}. For a system of such
particles there is the so called blocking temperature, $T_{\mathrm{B}}$,
below which the particle moment is blocked. The mean blocking temperature,
$\langle T_{\mathrm{B}}\rangle \propto K_{\mathrm{m}}V_{\mathrm{m}}$, is
determined by mean values ($K_{\mathrm{m}}$ and $V_{\mathrm{m}}$) of
the anisotropy constant and particle volume \cite{chant}. The above-mentioned
temperature $T_{\mathrm{g}}$ is directly related to
$\langle T_{\mathrm{B}}\rangle$ by relation
$T_{\mathrm{g}} =\beta\langle T_{\mathrm{B}}\rangle$ \cite{chant}, where
$\beta$ is a numerical factor of the order of unity depending on the form of
the particle size distribution. The higher is
temperature, the smaller is $H_{\mathrm{c}}$. According to Ref.~\cite{kat},
$H_{\mathrm{c}}\propto (1-T/T_{\mathrm{B}})^{1/2}$,
so that at $T>\langle T_{\mathrm{B}}\rangle $ the system becomes
superparamagnetic. It is therefore most probable that the linear
magnetic-field dependence of TKE in this film at $T=91$~K (near Kerr effect
maximum) (Fig.~3) is determined by the superparamagnetic behavior of the
clusters. The high $H_{\mathrm{p}}$ field for the film with
$x=0.15$ agrees with the $H_{\mathrm{c}}$ data obtained on a bulk polycrystal
of the same composition \cite{golovanov}.
\par
The FM interaction in clusters is mainly attributable to the
Co$^{3+}$(IS)-O-Co$^{4+}$(LS)  superexchange or the double-exchange of
localized $t_{2\mathrm{g}}$ electrons via itinerant $e_{\mathrm{g}}$
electrons \cite{senaris1}. As another source of FM state, the  magnetic
polarons formed near the Co$^{4+}$ ions can be mentioned \cite{yamaguchi1}.
The hole-rich regions stabilize IS state of Co$^{3+}$ ions at the interface
to the hole-poor regions. The clusters are thought to be coupled
antiferromagnetically (AFM) through the superexchange interaction between
the Co$^{3+}$ (IS) ions \cite{senaris1}.
Since the clusters are distributed randomly, the intercluster exchange is
frustrated and the cluster moments have non-collinear orientation.
This leads to a large anisotropy below the temperature $T_{\mathrm{g}}$.
The influence of all these factors on magnetic properties
can show itself as a typical spin-glass-like behavior.
\par
Among the films studied, this film has the highest MR [Fig. 6(c)]. It is
difficult to imagine that strengthening of the magnetic order in small
isolated clusters embedded in a semiconducting matrix, can cause this rather
strong effect \cite{belev}. It should be suggested, therefore, that an
applied magnetic field enhances the intercluster tunneling and/or
increases the conductivity of intercluster semiconducting regions.
It is quite probable that magnetic field affects somehow the interface
regions between hole-rich clusters and hole-poor matrix (which are enriched
with Co$^{3+}$ ions in IS state \cite{senaris1}). It was argued in Ref.
\onlinecite{caciuf} that the volume of the superparamagnetic clusters
should increase in a magnetic field, that can cause a considerable MR effect.
These suggestions need, however,
an elaboration and verification in a further study.

\subsubsection{$\mathrm{La}_{0.75}\mathrm{Sr}_{0.25}\mathrm{CoO}_{3}$ film}

When the Sr concentration is raised to $x=0.25$, the content of holes in
clusters increases. The volume of the FM clusters increases too and the
exchange in them is enhanced  by charge carriers (double exchange). A
decrease in transmittance with temperature decreasing on cooling below the
temperature of the metal-insulator transition $T_{\mathrm{MI}}\simeq 180$~K
[Fig. 10(a)] together with no signs of the transition to metallic state at
this temperature in $\rho (T)$ dependence (Fig.~4) indicates that in the
film with $x=0.25$ the transition to FM metallic state near 180 K occurs
only in the isolated hole-rich clusters (embedded in a semiconducting
matrix) rather than in the whole volume of the film. For this reason there
are no continuous metallic conducting paths penetrating the whole film and,
therefore, the transition of the clusters to the FM state has no visible
effect on recorded $\rho (T)$ behavior (Fig.~4). But this transition is
reflected in the MR temperature behavior which is shown in Fig.~6(b). It can
be
seen that the MR reaches its maximum below $T\simeq 200$~K which agrees well
with the magneto-optical and optical data obtained [see Figs.~1(b) and
10(a)]. Generally, the temperature behavior of MR for this film corresponds
to that of highly inhomogeneous magnetic system, as it was mentioned
already in Sec.~\ref{resist}.
\par
No significant MR anisotropy or MR hysteresis has been found in this film.
It is caused, perhaps, by proximity of composition to the
percolation threshold ($x_{\mathrm{c}} =$ 0.20--0.25), near which cluster sizes and
orientations are most chaotic.  TKE appears at $T\simeq 230$~K, i.e. above
the temperature $T_{\mathrm{MI}}\simeq 180$~K  obtained from the
transmittance data. The relation $T_{\mathrm{MI}} < T_{\mathrm{c}}$ is
expected for this range of Sr doping \cite{caciuf}. The
field dependence of the TKE (Fig.~3) exhibits a typical FM behavior at the
temperature ($T\approx 180$~K), where the TKE is maximum, but it is linear
at $T=70$~K, which is typical of superparamagnetic system, but can also
be determined by a change in the relationship between the FM and AFM
contributions responsible for magnetism of this film with lowering
temperature.

\subsubsection{$\mathrm{La}_{0.65}\mathrm{Sr}_{0.35}\mathrm{CoO}_{3}$ film}

For $x>x_{\mathrm{c}}$, the volume fraction of hole-rich regions should be
high enough to form an infinite percolating FM cluster below $T_{\mathrm{c}}$
\cite{senaris1}. The results for the film with $x=0.35$
support well this view. Both, the $\rho (T)$ and $I(T)$ dependences
[Figs. 5 and 10(a)] manifest that the metal-insulator transition starts at
$T\approx 250$~K. The temperature behavior of the MR is typical of
optimally doped cobaltites, with a sharp maximum at $T=230$~K which
corresponds to the Curie temperature for this film. At that, the metallic
behavior of $I(T)$ is more pronounced than for $x=0.25$, but a strong
anomaly can be seen in the range 160--220 K. The maximum Kerr effect is
found to be larger for $x=0.35$ than that of for $x=0.25$, and the
temperature range of maximal Kerr effect magnitude is wider [Fig. 1(c)].
At the same time the magnitude of Kerr effect drops sharply below 160 K.
The same as for $x=0.25$, the field dependence of TKE (Fig. 3)
exhibits a typical FM behavior for the temperature range, where
$\delta(T)$ is maximum, but it is linear at $T=96$~K.
\par
Although at $x=0.35$ the infinite FM cluster, percolating through the whole
sample, is formed, at the same time, however, a pervasive hole-poor matrix
with some isolated clusters in it persists up to $x=0.5$ \cite{senaris1}.
The large cluster interfaces and/or the interlayer between the FM grains
contain Co$^{3+}$ (IS) ions which change into the LS state with lowering
temperature and no longer assist the charge transport and Kerr effect. As
the number of Co$^{3+}$ (IS) ions reduces and the related FM interaction
grows weaker, the carriers  can gradually localize in the FM clusters
themselves. This mechanism is maybe responsible for the rather high
resistivity of the film with $x=0.35$ below $T=50$~K and for the hysteretic
and anisotropic behavior of MR. The low-temperature resistance minimum is
typical of polycrystalline samples with weak enough interconnections between
FM grains. In this case the intergrain tunneling can become activated at low
enough temperature that leads to non-metallic behavior of $\rho(T)$
\cite{belev}. It is clear that magnetical inhomogeneity due to the phase
separation can contribute to this effect in the same way as
polycrystalline structure. The MR hysteresis is found to be absent
above 90 K, but fields $H_{\mathrm{p}}$ are significant already at $T=78$~K
($H_{\mathrm{p}\parallel}=2.5$~kOe and $H_{\mathrm{p}\perp}=2.0$~kOe).
With temperature decreasing down to $T=4.2$~K the $H_{\mathrm{p}}$ values
have increased up to $H_{\mathrm{p}\parallel}=4.5$~kOe and
$H_{\mathrm{p}\perp}=5.3$~kOe. The strong decrease in TKE and change in
character of field dependence of it for decreasing temperature, together with
rather high MR anisotropy and the considerable growth of $H_{\mathrm{p}}$
in low temperature range, indicate that in spite of metallic behavior of
the film conductivity, FM state in it is not homogeneous and maybe close to
cluster-glass state.

\subsection{Transmittance anomaly and orbital ordering of Co$^{3+}$
(IS) ions.}
The features of the temperature and spectral behavior of
optical and magneto-optical properties, considered above, can be
determined by the transition of Co$^{3+}$ ions from LS state to
either or both HS and IS states. In the transmittance spectra, however,
the special anomaly is found, which can be attributed only to the
LS--IS transition. This anomaly in the $I(T)$ dependence (kinked behavior of
it) is observed for all films studied and appears practically in the same
temperature range 160--220 K [Fig.~10(a)]. It can be seen quite clearly for
the films with $x=0.15$ and $x=0.35$ on the background of the general
semiconductive or metallic run of their $I(T)$ curves. For the film with
$x=0.25$ the temperature range of the anomaly nearly coincides with the
temperature $T_{\mathrm{MI}}$ and, therefore, the anomaly is not so distinct.
The anomaly of this type is never mentioned for manganites. It is
reasonable, therefore, to assume that its nature is determined by some
specific feature of cobaltites. This is the transition of Co$^{3+}$ ions
from LS state to a higher-spin state with increasing temperature. According
to the Ising model molecular-field calculations, an energy gap,
$\Delta =230$~K, between the $S=0$ and $S=1$ spin states provides a good
description for the temperature dependence of magnetic susceptibility
\cite{tokura}. The spin transition proceeds gradually. No information can be
found in literature about special temperature points of this transition,
except for the M\"{o}ssbauer effect in LaCoO$_3$ \cite{bhide}. The latter
data permit the conclusion that the ratio between the Co$^{3+}$ (HS)
concentration and the total Co$^{3+}$ content reaches its maximum at
$T\simeq 200$~K. It follows, however, from neutron scattering data for
LaCoO$_3$ \cite{caciuf} that the 50:50 ratio of low-spin and higher-spin
Co$^{3+}$ ions is stable in a wide interval from 110~K to room temperature,
which is supported by the results for magnetic susceptibility
\cite{senaris2}.
\par
According to electron structure calculation for LaCoO$_3$ \cite{korotin},
with rising temperature the LS ($S=0$) state of Co ions transfer to the
IS ($S=1$) state rather than to HS ($S=2$) one. Energetically, the latter
state appears to be higher than the IS state even above $T\simeq 600$~K.
The Co$^{3+}$ (IS) ion with the ($t_{2\mathrm{g}}^{5}e_{\mathrm{g}}^{1}$)
configuration is a Jahn-Teller (JT) ion. In the case of equal spins, the
orbitally ordered (non-metallic) state is more preferable than the state
without orbital order \cite{korotin}. In this case the non-metallic behavior
of LaCoO$_3$ at $90$~K~$<T<500$~K can be described properly assuming that
Co$^{3+}$ ions are in the orbital-ordered state. The gradual
transition to the metallic behavior observed experimentally at $T>550$~K is
associated with the destroying the orbital order \cite{korotin}.
The presence of JT Co$^{3+}$ (IS) ions in cobaltites
is supported by experimental results, such as anomalous splitting of phonon
modes in LaCoO$_3$ \cite{yamaguchi3} and giant anisotropic magnetostriction
in La$_{1-x}$Sr$_{x}$CoO$_3$ \cite{ibarra}. Measurement of magnetic circular
dichroism spectra has revealed the presence of Co$^{3+}$ (IS) ions with a
finite orbital moment \cite{yoshi}.
\par
It is possible that the $I(T)$ anomaly appears because the
orbital-ordered Co$^{3+}$ (IS) ions reach their maximum concentration at
$T\simeq 180$~K. At $T>180$~K the ions transfer gradually into the
orbital-disordered state. The anomalous splitting of phonon modes in the
spectrum of LaCoO$_{3}$ \cite{yamaguchi3} strengthens this hypothesis.
The mode corresponding to the orbital-ordered state becomes saturated at
$T\simeq 200$~K whereas the intensity of the mode ascribed to the
orbital-disordered state increases sharply above $T\simeq 160$~K.
\par
As the LS-IS transition occurs in some fraction of the Co ions, two
features are expected in the optical spectra of mixed-valence cobaltites:
(i) appearance of new absorption lines corresponding to the optical
transitions including energy states of Co (IS) ions and
(ii) charge-carrier localization caused by JT lattice distortions near
Co$^{3+}$ (IS) ions, which may entail the formation of JT polarons.
 The intricate $I(T)$
dependence near the absorption edge (1.0--1.4 eV) [Fig.~10(b)]
is determined, probably, by the appearance of additional absorption
bands, which, however, are not resolved in this spectral range.
These bands can have different temperature dependence,
including the anomalous ones, and their superposition would result in
tangled appearance of $I(T)$ dependence.
\par
A possible reason for the $I(T)$ anomaly in the region of light
interaction with the charge carriers at $E=0.20$~eV [Fig. 10(a)] can be
charge localization at the maximum concentration of the orbital-ordered
JT Co$^{3+}$ (IS) ions. It is natural to assume that the orbital-ordered
Co$^{3+}$ (IS) ions reside in the hole-poor matrix (for $x=0.15$ and 0.25)
or in the regions near the boundaries of hole-rich clusters (for $x=0.35$).
The hole-rich clusters are expected to contain orbital-disordered
Co$^{3+}$ (IS) ions. The $I(T)$ anomaly can appear only when contributions
of these two mechanisms -- enhancing of metallic behavior of charge carriers
in the clusters and localization of charge carriers in the matrix -- are
comparable. It should be noted that this mechanism of charge-carrier
localization is possible only for JT ions in semiconducting matrix, that is
for Co (IS) ions. There is no such mechanism for LS-HS transition of
Co$^{3+}$ ions. Cooling in a finite applied field, or cooling at $H=0$ and
subsequent measurement at $H>0$, suppresses the anomaly in $I(T)$, i.~e.
enhances the metallic contribution of the charge carriers [see Fig.~10(c) for
$x=0.35$].  The anomaly is affected most significantly by fast cooling of
the film preheated to $T=320$~K [Fig.~10(c)]. At this temperature the
significant fraction of Co$^{3+}$ (IS) ions in the semiconducting matrix
becomes orbital-disordered. On fast cooling this "disordered" (and
more metallic) state is frozen. After such thermal treatment the $I(T)$
anomaly vanishes completely both at $E=0.20$~eV (Fig.~10) and at
$E=1.0$~eV (not shown). The metallic contribution becomes more  pronounced
as well. Appearance and change in number of orbital-ordered Co$^{3+}$ (IS)
ions in dielectric matrix can be responsible for unusual behavior of TKE
temperature dependence for films with $x\geq 0.25$ as well.
\par
At $T=80$~K,  Kerr effect is equally small in the films with $x=0.25$ and
$x=0.35$, and it is smaller than in the film
with $x=0.15$ (Fig.~1). This is because only a small fraction of Co$^{3+}$
(IS) ions are involved in the FM exchange at 70--90 K. In films with
$x\geq 0.25$, Co$^{3+}$ (IS) ions are fewer than in the film with $x=0.15$
if we assume that the highest Co$^{3+}$ (IS) concentration is 50\% of all
Co$^{3+}$ ions both in non-doped LaCoO$_3$ and doped
La$_{1-x}$Sr$_{x}$CoO$_{3}$.  Besides, it is likely that in the matrix and
in the boundary regions of metallic clusters the Co$^{3+}$ (IS)
ions are in the orbital-ordered state, so they interact
antiferromagnetically. As a result, the number of the ions
involved in the FM interaction is much less than 50\% of the total
Co$^{3+}$ content corresponding to the nominal film composition. The linear
behavior of TKE as a function of magnetic field which is observed in all
the films below the TKE peak (Fig.~3) shows that the FM interaction (double
exchange) weakens, and concurrence the FM interaction (in clusters) and AFM
interaction (in the matrix or boundary regions) comes into play.
\par
An unusual fact is observed for the film with $x=0.35$: on heating in
a weak field ($H=0.9$~ kOe), the TKE dependence shifts significantly
towards higher temperature when compared with that of at $H=3.5$~kOe
[Fig.~1(c)]. This finding implies that when Kerr effect is measured in the
magnetic field 3.5 kOe, the
Co$^{3+}$ (LS)-Co$^{3+}$ (IS) transition is driven not only by temperature
but by the magnetic field as well. The spin transition should actually occur
at higher temperatures, and the magnetic field stimulates its onset at lower
temperatures. This conclusion was arrived in Ref. \onlinecite{ibarra}
at explaining the giant anisotropic magnetostriction of
La$_{1-x}$Sr$_{x}$CoO$_3$ ($x=0.3$).
The  curves obtained for the film with $x=0.35$ are shifted after
heating-cooling in a strong field [Fig.~3(c)] because heating stimulates
an earlier onset of the LS-IS transition and cooling "freezes" the state
which existed at the high temperature. The temperature hysteresis of TKE
is small at $x=0.25$ presumably because the onset of FM and metal-insulator
transition in clusters  takes place within the
interval $T=$180--200 K, where the content of orbital-ordered JT ions and,
hence, the related AFM contribution are the highest. The state without
orbital order has a larger magnetic moment \cite{korotin}, which affords
a strong TKE even above the temperature of metal-insulator transition. The
considerable TKE shift towards higher
temperatures caused  by fast cooling of the film with $x=0.25$ [Fig.~1(b)]
agrees well with the effect produced by fast cooling on the $I(T)$ curves
[Fig.~10(c)]. The fast cooling of the film which is partially in the
orbital-disordered state (at $T\geq 300$~K) is favorable for "freezing" this state.
As a result, the carrier localization weakens, the AFM component decreases
and the FM contribution increases.
\par
The evolution of optical, magneto-optical and transport properties of the
films studied with the doping level correlates well with the concept of
electron phase separation in cobaltites. At a low doping level ($x=0.15$),
the Co$^{3+}$ (LS)-Co$^{3+}$ (IS) transition in the magnetic field begins
at helium temperatures. The transition enhances the FM contribution in the
clusters, which is mainly connected with the Co$^{3+}$ (IS) - O -Co$^{4+}$
(LS) superexchange. As a result, the Kerr effect grows up to $T=118$~K. On
the farther rise of temperature the number of Co$^{3+}$ (IS) ions increases. The
content of orbital-ordered ions in the matrix increases too and so does the
related AFM contribution. In the temperature region, where the content of
the orbital-ordered Co$^{3+}$ (IS) ions is the highest ($T\simeq 180$~K),
the AFM contribution exceeds the FM one from the clusters and the Kerr
effect turns to zero. The high coercive forces are indicative of small sizes
of the FM clusters in the non-magnetic matrix. The magnetic state of the
film with $x=0.15$ has doubtless features of the spin-glass state.
\par
On doping level ($x=0.25$) near the percolation threshold, the hole-rich FM
clusters are quite numerous, but segregated. No infinite cluster is formed
yet.  In the clusters the metal-insulator transition occurs near 180 K. This
suggest that below 180 K the FM interaction is mainly connected
with the charge carrier exchange (double exchange). Because of doping,
the total content of Co$^{3+}$ ions is smaller than at $x=0.15$ and the Kerr
effect is also smaller at low temperatures. It is likely that the fraction
of the orbital-ordered Co$^{3+}$ (IS) ions grows with temperature only in
the matrix and becomes the highest at $T\simeq 180$~K. Since the degree of
orbital ordering decreases above this temperature, TKE remains appreciably
strong up to $T=230$~K because the state without orbital ordering has a
large moment.
\par
At the doping level ($x=0.35$) above the percolation threshold, the FM
regions percolate magnetically as well as conductively through the the whole
film. For this reason the temperature of metal-insulator transition is close
to $T_{\mathrm{c}}= 230$~K. Although the number of Co$^{3+}$ ions involved
in the LS-IS transition decreased, the TKE maximum is much higher than at
$x=0.25$ because the double exchange assisted by Co$^{4+}$ ions enhances the
FM contribution. The film is, however, magnetically inhomogeneous, which is
evident from the TKE hysteresis under the heating-cooling condition. The
considerable hysteresis and the strong dependence of TKE on the sample
prehistory and thermal treatment may be connected with orbital ordering in
the cluster interfaces, which are the sources of the AFM contribution.
As in the $x=0.25$ case, the FM state of the film with $x=0.35$
below the temperature of TKE maximum becomes much weaker due to the enhanced
AFM interaction. This is evident from the sharp growth of $H_{\mathrm{c}}$
with decreasing temperature  typical of the cluster glass state.

\section{Conclusion}
Optical, magneto-optical and transport properties of polycrystalline
La$_{1-x}$Sr$_{x}$CoO$_3$ ($x=0.15, 0.25, 0.35$) films have been studied.
Unlike manganites, no direct correlation is found between the temperature of
metal-insulator transition in clusters and the temperature at which the
FM contribution appears. It is found that the temperature dependences of the
optical and magneto-optical properties of the films exhibit features, which
can be determined by the transition of Co$^{3+}$ ions from the low-spin
state ($S=0$) to the intermediate spin state $S=1$. This transition is driven
not only by temperature increase but by an applied magnetic field as well.
It is shown as well that the above-indicated properties
are dependent on the number of orbital-ordered Co$^{3+}$ ions in
semiconducting matrix.
Irrespective of the Sr concentration, the content of the orbital-ordered
Co$^{3+}$ ions is the highest at $T\simeq 180$~K. The results obtained
support the phase separation scenario for mixed-valence cobaltites.

\begin{acknowledgments}
The authors thank M. A. Korotin for helpful discussions. The study is
partially supported by RFBR Grants No. 02-02-16429 and No. 00-02-17797.
\end{acknowledgments}


\newpage
\centerline{{\bf Figure captions}}
\vspace{15pt}
\par
Figure 1. Temperature dependences of TKE of
La$_{1-x}$Sr$_{x}$CoO$_3$ films with $x=0.15$ (a), $x=0.25$ (b) and
$x=0.35$ (c). Heating and cooling regimes are indicated by arrows.
All curves were recorded in magnetic field $H=3.5$~kOe at photon energy
$E=2.8$~eV, except dashed-line curve for $x=0.35$, which was recorded
on heating in field $H=0.9$~kOe. Solid line in the picture for
$x=0.25$ (b) corresponds to fast cooling. Rates of heating
and cooling were usually in the range 1--3 K/min, the ``fast cooling''
was done with a rate about 20 K/min.
\vspace{14pt}
\par
Figure 2. Spectral dependences of TKE of the La$_{1-x}$Sr$_{x}$CoO$_3$
films at temperatures of maximum in the temperature
dependence of TKE.
\vspace{14pt}
\par
Figure 3. Magnetic-field dependences of TKE of the
La$_{1-x}$Sr$_{x}$CoO$_3$ films studied at temperatures,
corresponding to the maximal TKE effect (190 K for $x=0.35$ and
180 K for $x=0.25$), and at much lower temperatures.
\vspace{14pt}
\par
Figure 4. Temperature dependences of the resistivity
of the films La$_{1-x}$Sr$_{x}$CoO$_3$ (in
semilogarithmic coordinates).
\vspace{14pt}
\par
Figure 5. Temperature dependence of the resistivity
of film La$_{0.65}$Sr$_{0.35}$CoO$_3$.
\vspace{14pt}
\par
Figure 6. Temperature dependences of magnetoresitance
$[R(H)-R(0)]/R(0)=\Delta R(H)/R(0)$ at $H=20$~kOe of the
La$_{1-x}$Sr$_{x}$CoO$_3$ films. The fields $H_{\parallel}$
and $H_{\perp}$ were  applied parallel and perpendicular to the film plane,
respectively. In both cases the fields  were perpendicular to
the transport current. The MR of the film with $x=0.25$ (b) has not shown a
noticeable sensitivity to the angle between the field and the film plane.
\vspace{14pt}
\par
Figure 7. Magnetoresistive hysteresis for  La$_{1-x}$Sr$_{x}$CoO$_3$
film with $x=0.35$ at $T=78$~K. Applied fields were parallel
($H_{\parallel}$) and perpendicular ($H_{\perp}$) to the film plane.
In both cases the fields were perpendicular to the transport current.
\vspace{14pt}
\par
Figure 8. Optical density spectra $D=\ln(I_{0}/I)$ ($I_0$ is the
incident-light intensity, $I$ is the intensity of light transmitted through
the film) for La$_{1-x}$Sr$_{x}$CoO$_3$ films studied in the visible light
range at $T=295$~K. The curves are spaced apart for facilitation of
visual perception.
\vspace{14pt}
\par
Figure 9. Optical density spectra $D=\ln(I_{0}/I)$ ($I_0$ is the
incident-light intensity, $I$ is the intensity of light transmitted through
the film) for La$_{1-x}$Sr$_{x}$CoO$_3$ films studied in the IR range at
different temperatures.
\vspace{14pt}
\par
Figure 10. Temperature dependences of the intensity of light transmitted
through the La$_{1-x}$Sr$_{x}$CoO$_3$ films at $E=0.20$~eV (a), in the energy
range 1.0--1.4 eV (b) and for different measurement conditions at $E=0.20$~eV
for film with $x=0.35$ (c). The numerals in panel (c) of the figure
indicate the following: 1 -- ZFC, $H=0$; 2 -- FC, $H=0$; 3 -- ZFC,
$H=8$~kOe; 4 -- fast cooling after a heating up to $T=320$~K, $H=0$
(heating with the rate about 1--1.5 K/min; the following "fast cooling"
from 320 K  to $T\approx 80$~K have taken about 10 min). The curves in
panels (a) and (b) are spaced apart for facilitation of visual perception.
}
\newpage

\begin{figure}
\centerline{\epsfig{file=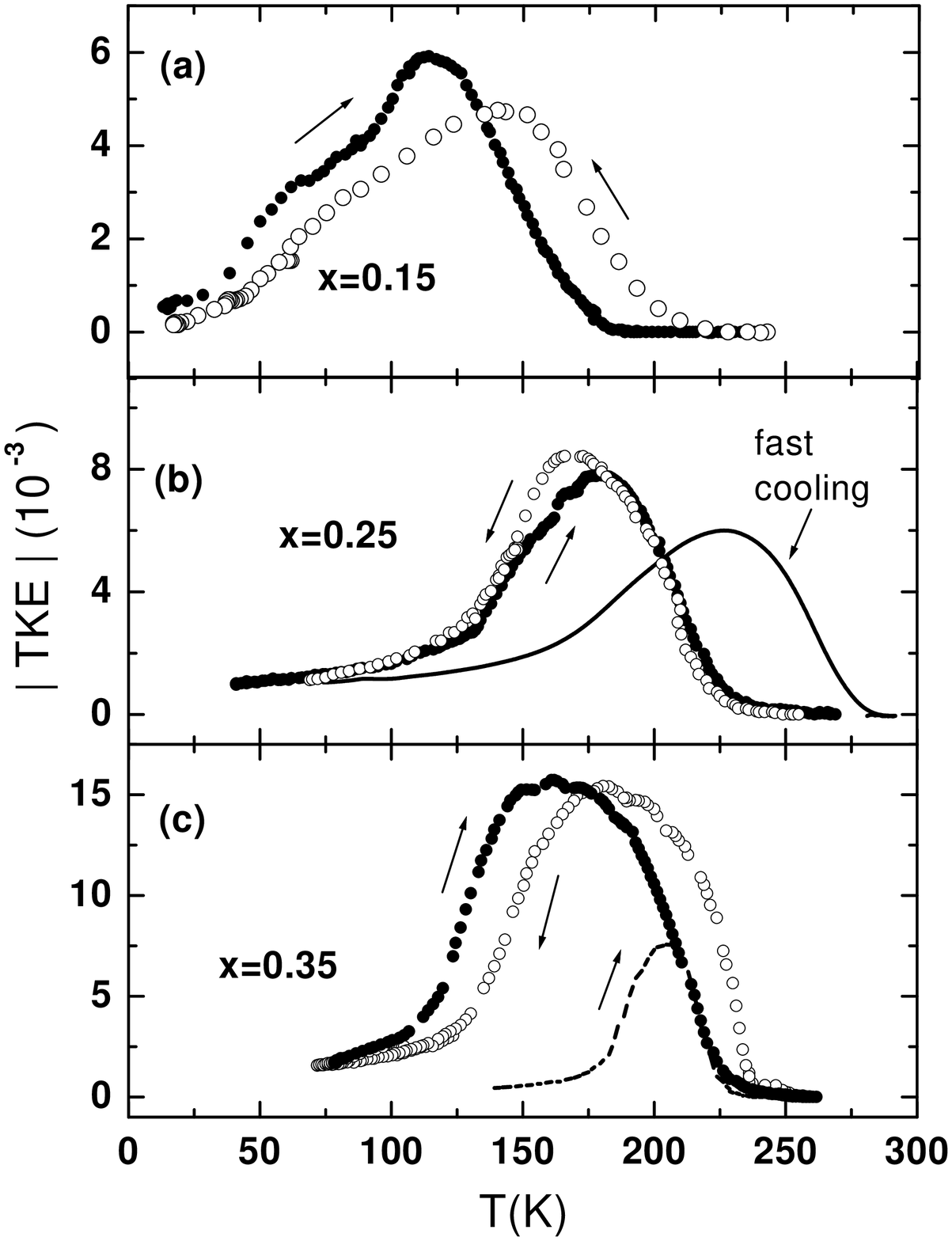,width=12cm}}
\vspace{15pt}
Figure 1 to paper Loshkareva et al. (Phys. Rev. B)
\end{figure}

\newpage

\begin{figure}
\centerline{\epsfig{file=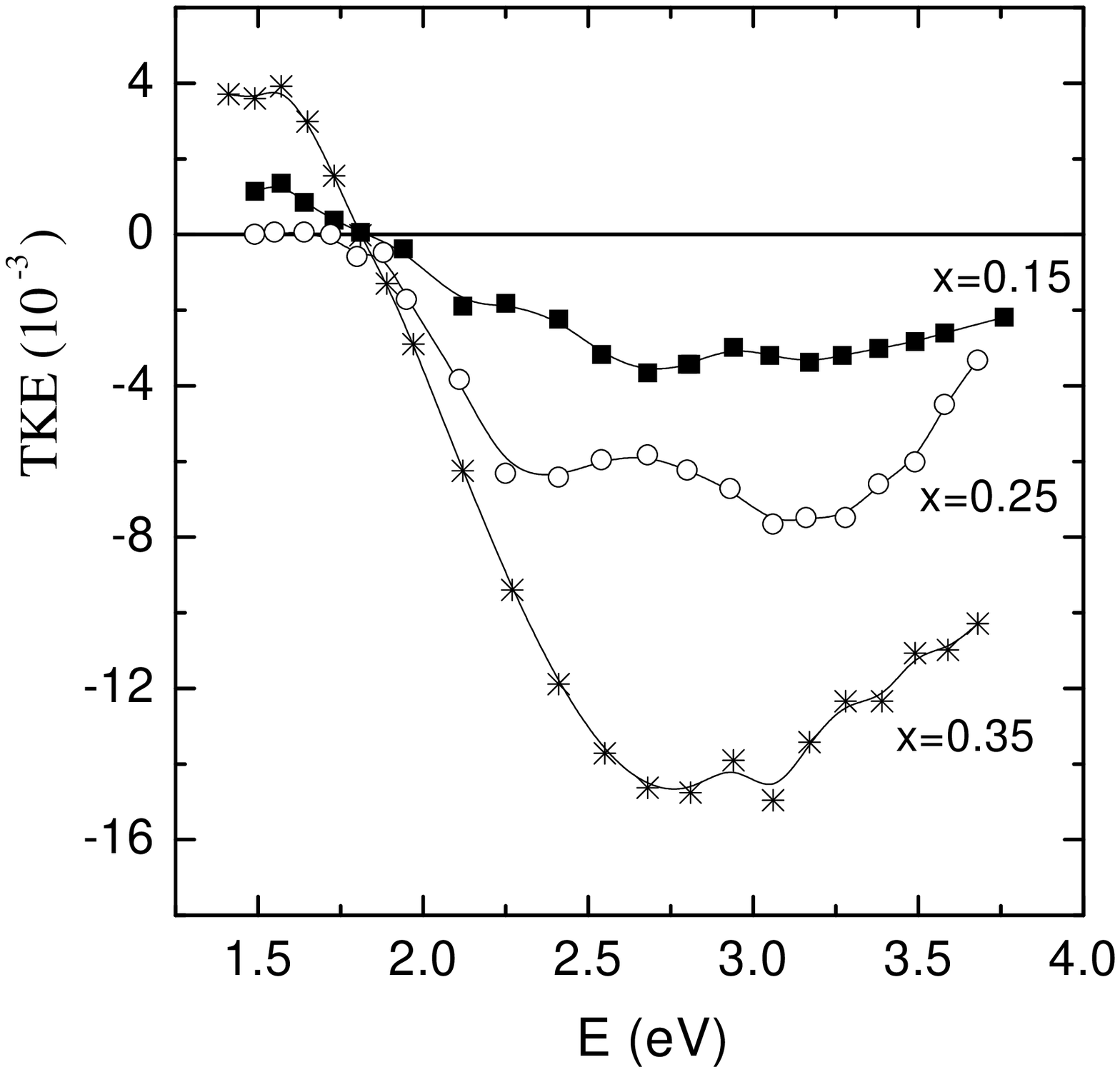,width=10.5cm}}
\vspace{12pt}
Figure 2 to paper Loshkareva et al. (Phys. Rev. B)
\end{figure}

\begin{figure}
\centerline{\epsfig{file=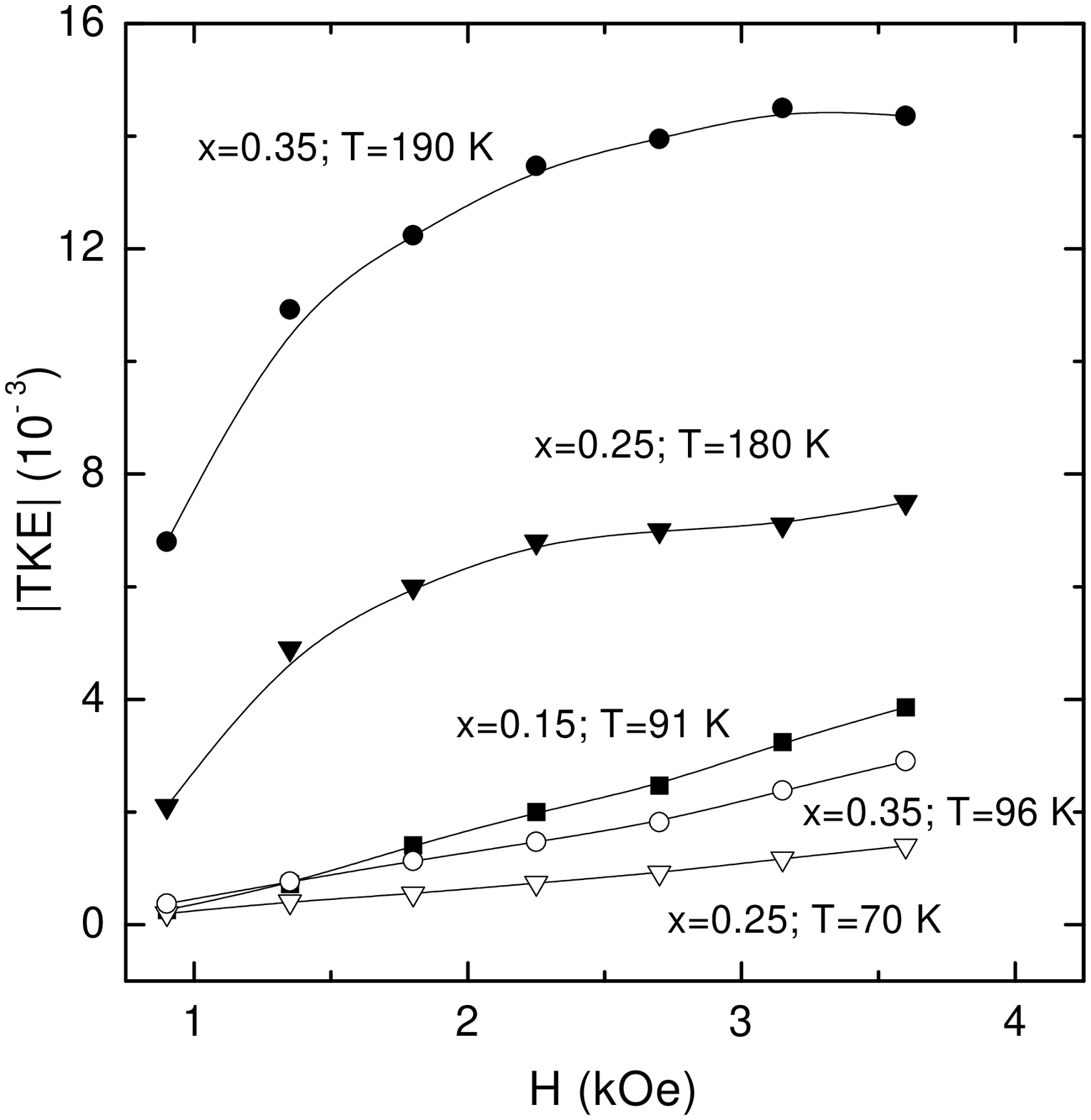,width=10cm}}
\vspace{12pt}
Figure 3 to paper Loshkareva et al. (Phys. Rev. B)
\end{figure}

\newpage
\begin{figure}
\centerline{\epsfig{file=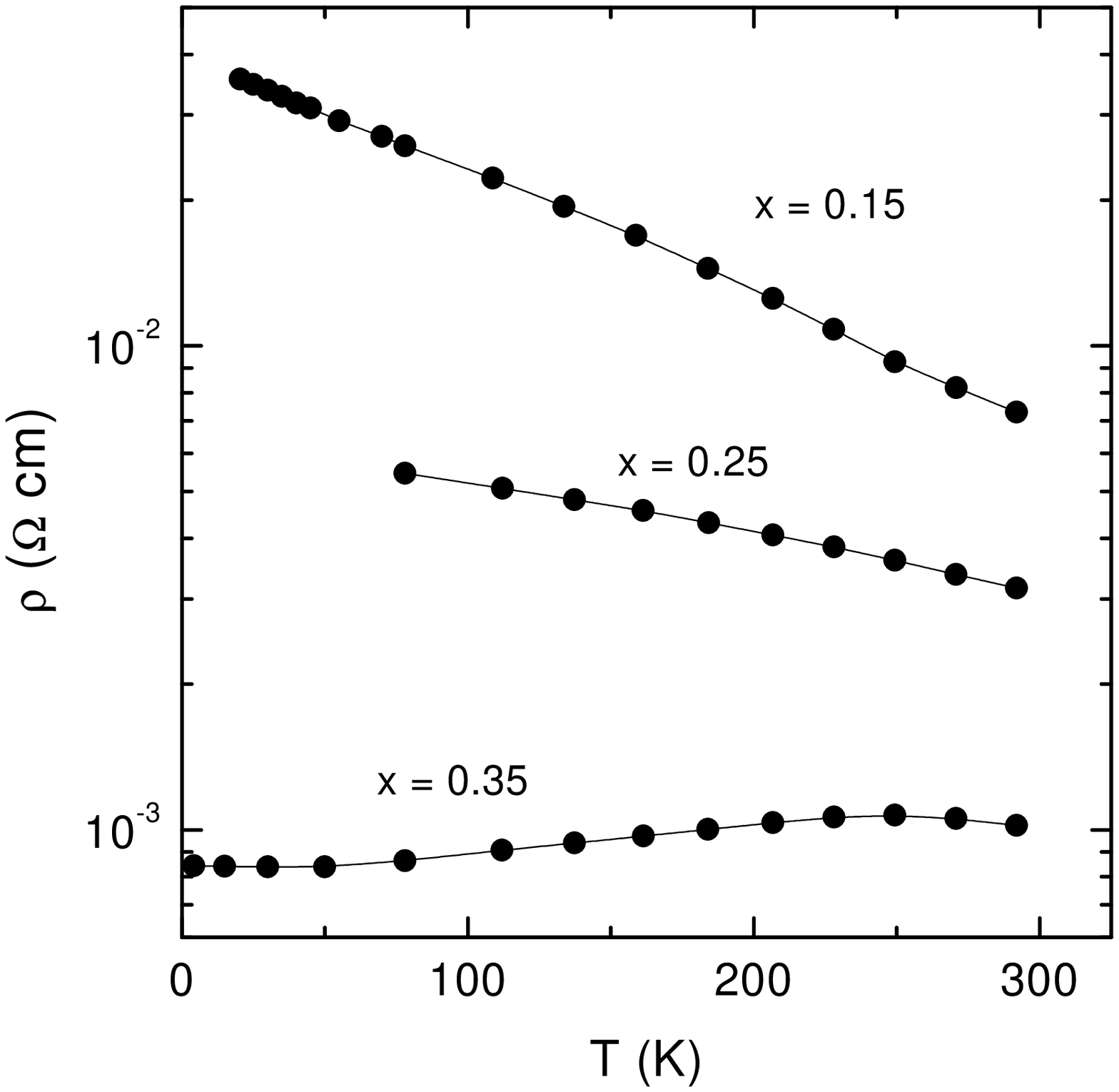,width=10cm}}
\vspace{12pt}
Figure 4 to paper Loshkareva et al. (Phys. Rev. B)
\end{figure}

\begin{figure}
\centerline{\epsfig{file=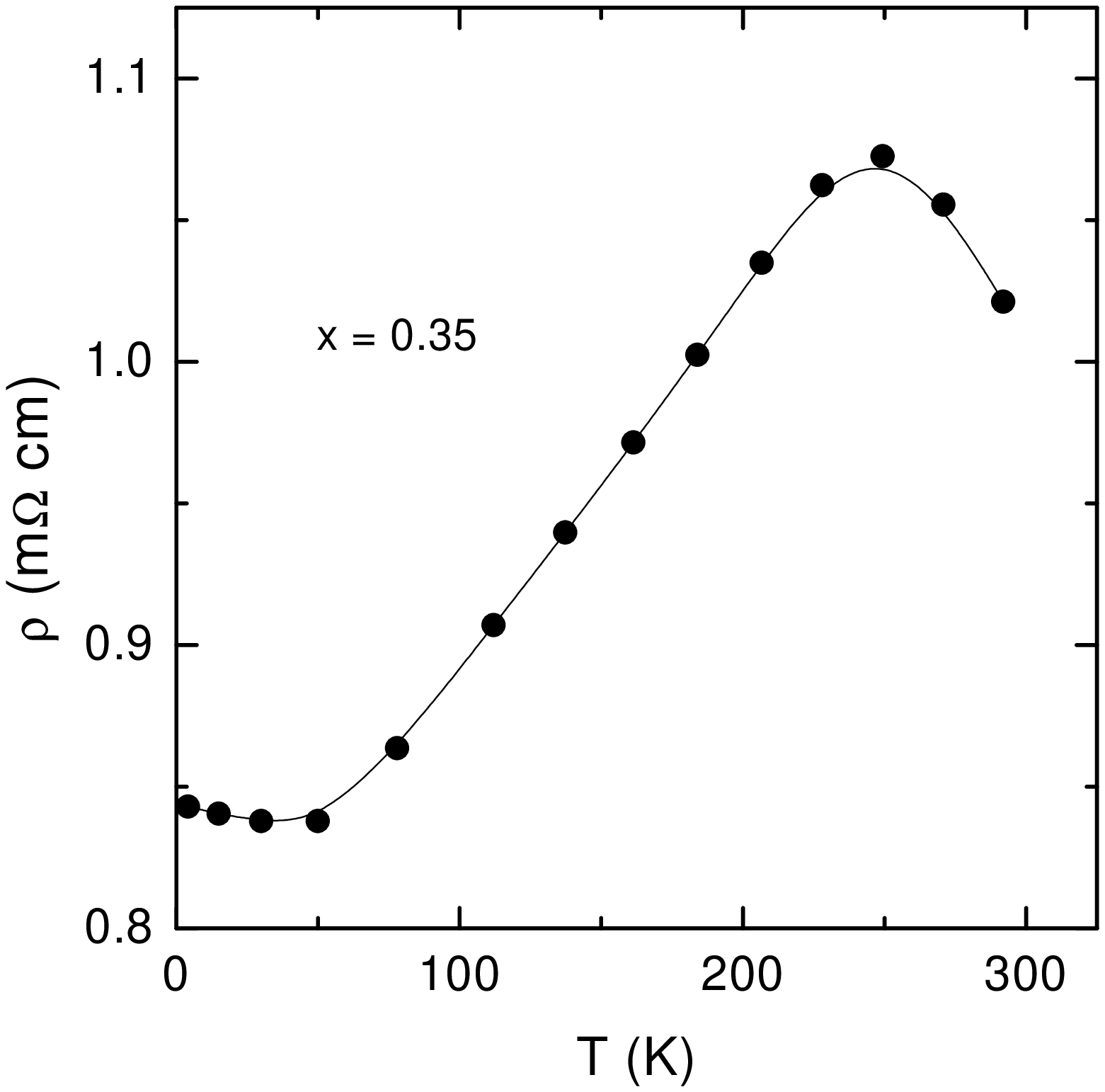,width=10.5cm}}
\vspace{12pt}
Figure 5 to paper Loshkareva et al. (Phys. Rev. B)
\end{figure}

\newpage
\begin{figure}
\centerline{\epsfig{file=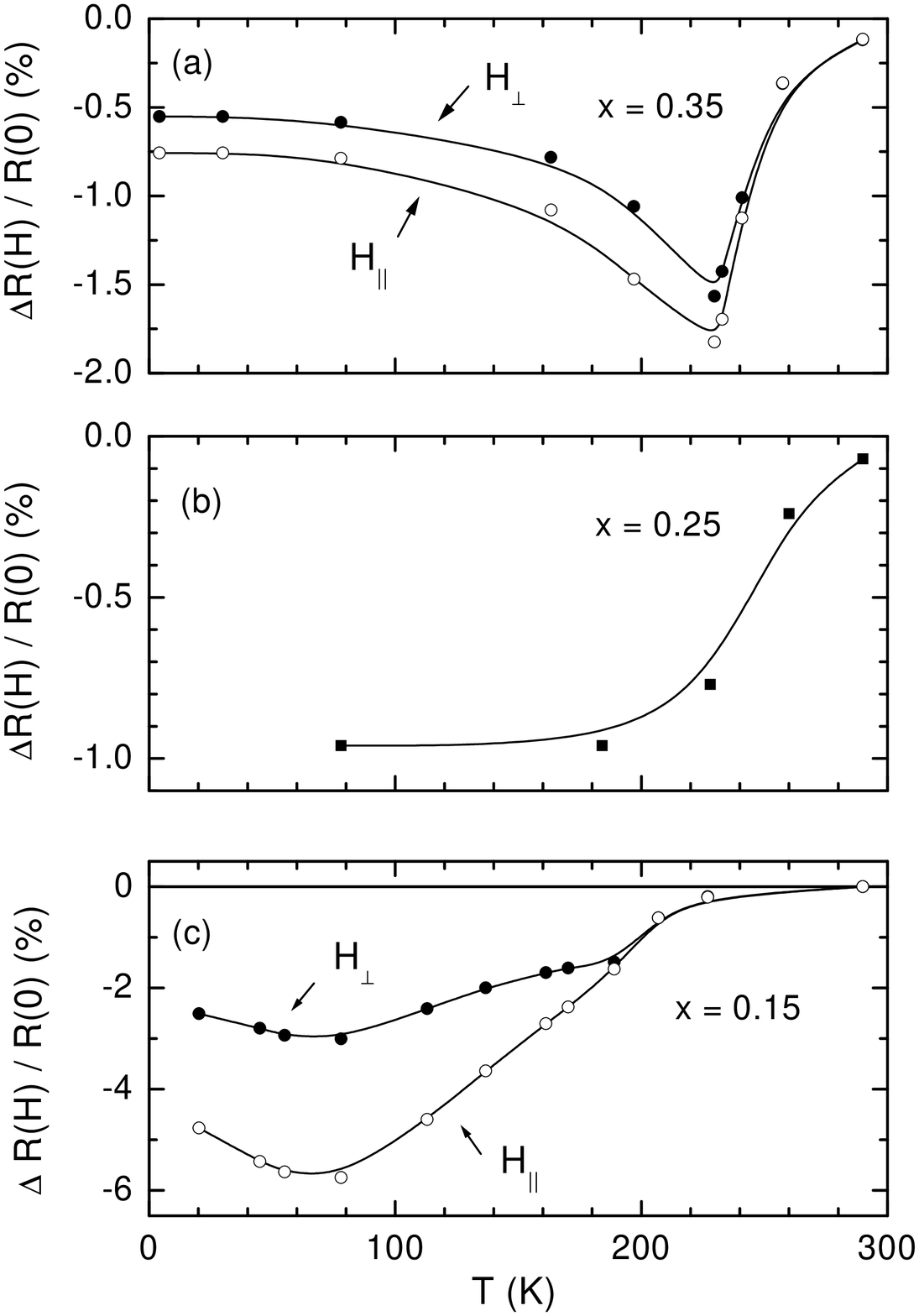,width=12cm}}
\vspace{25pt}
Figure 6 to paper Loshkareva et al. (Phys. Rev. B)
\end{figure}

\newpage
\begin{figure}
\centerline{\epsfig{file=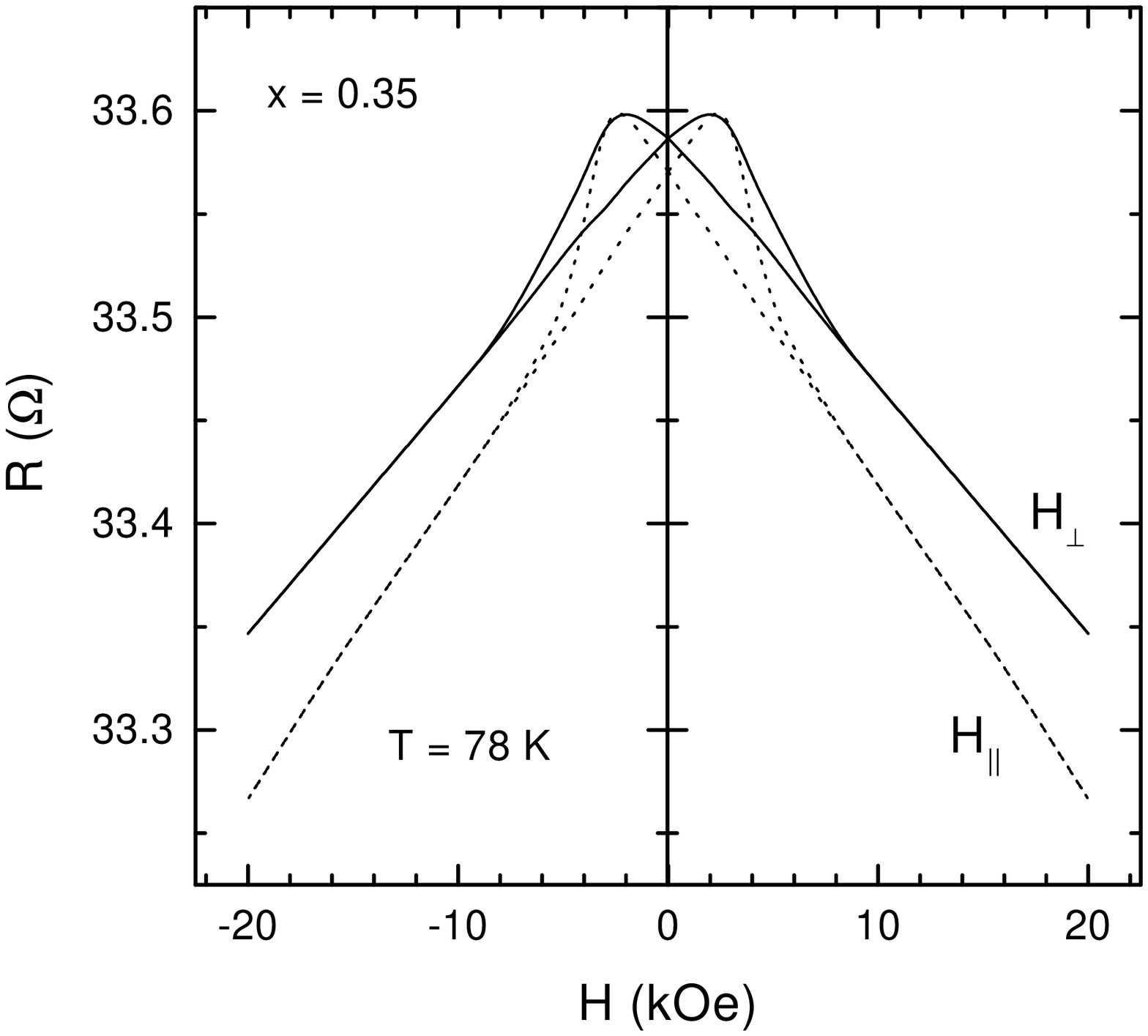,width=10.5cm}}
\vspace{12pt}
Figure 7 to paper Loshkareva et al. (Phys. Rev. B)
\end{figure}

\begin{figure}
\centerline{\epsfig{file=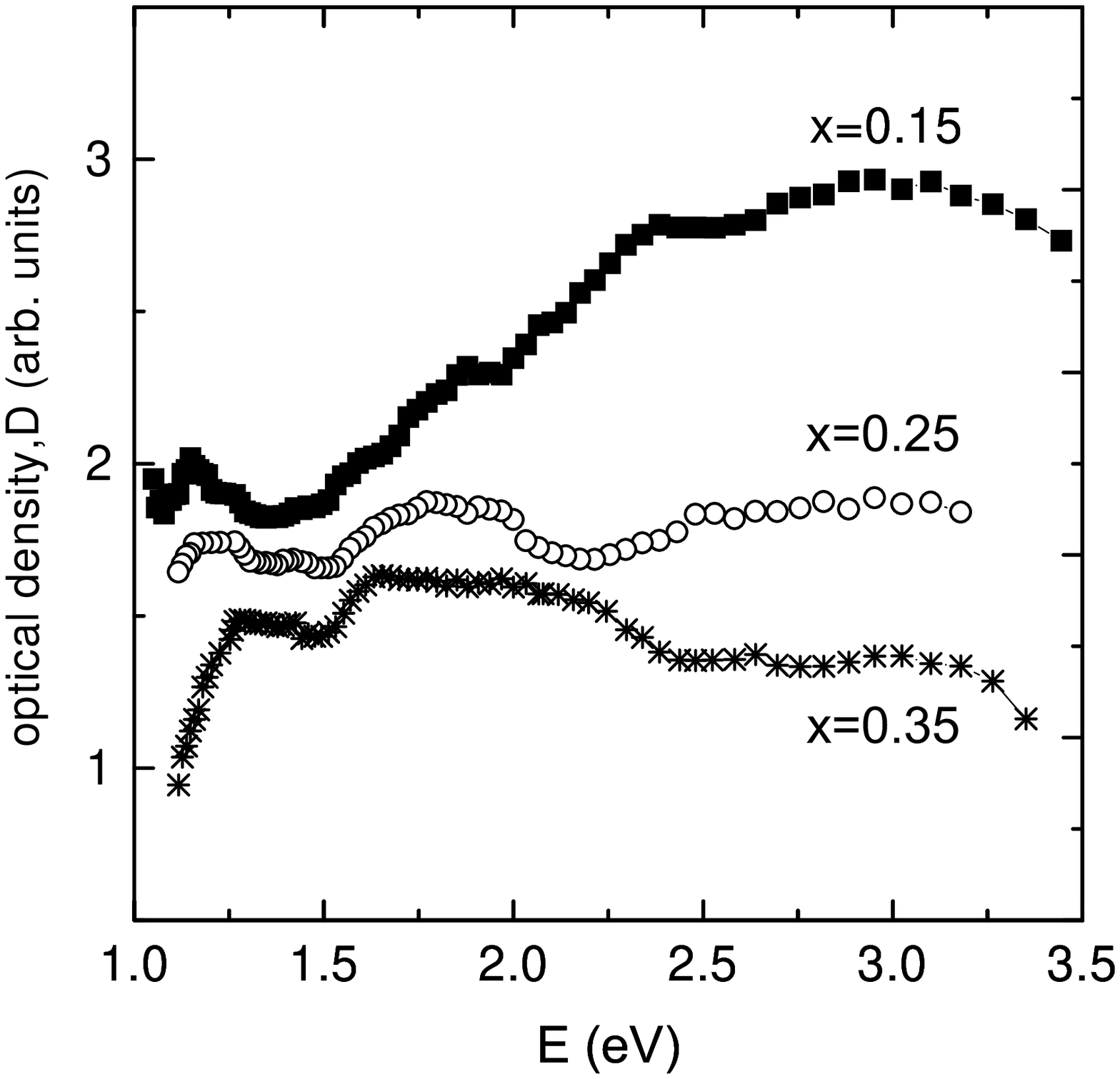,width=10cm}}
\vspace{12pt}
Figure 8 to paper Loshkareva et al. (Phys. Rev. B)
\end{figure}

\newpage
\begin{figure}
\centerline{\epsfig{file=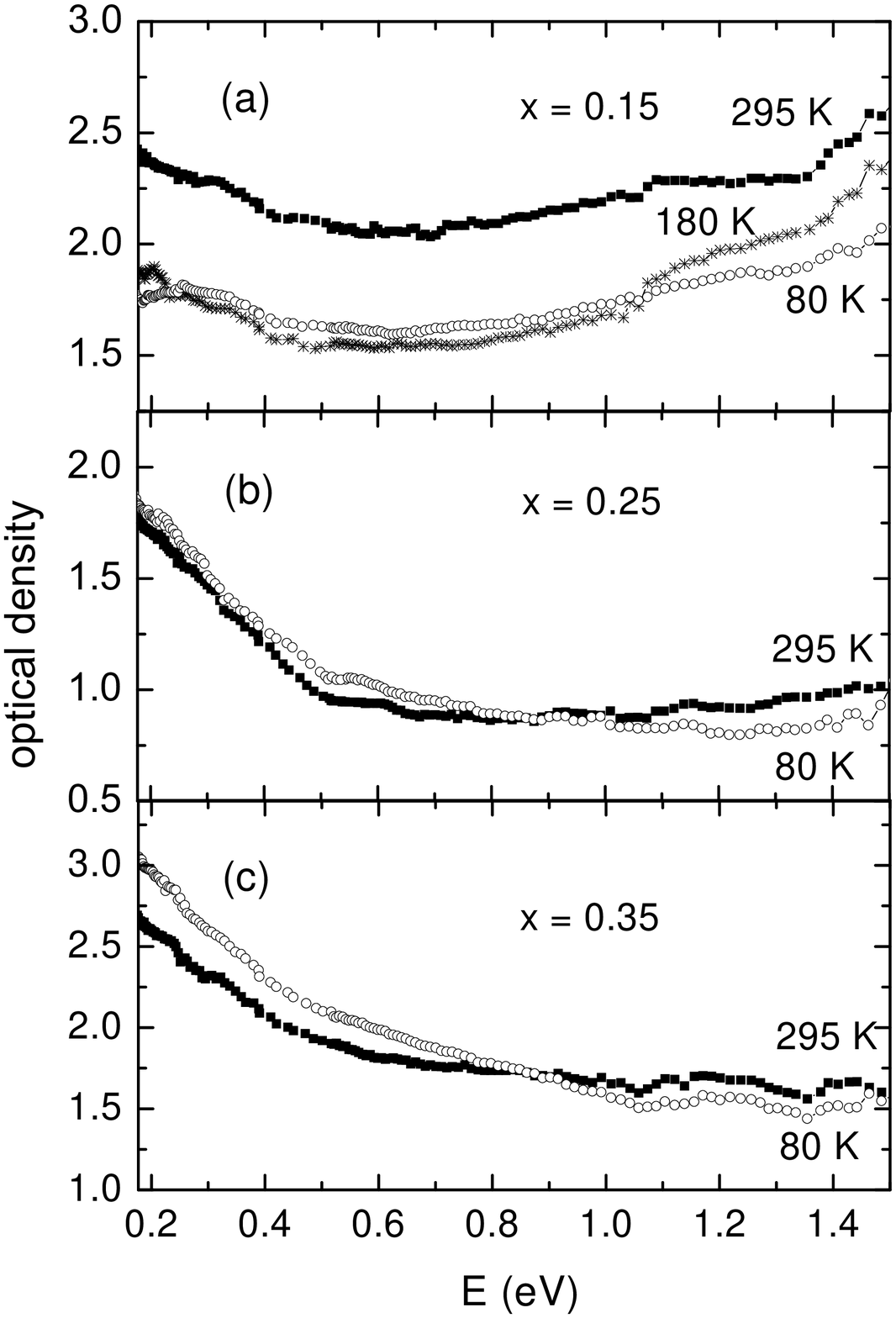,width=12cm}}
\vspace{25pt}
Figure 9 to paper Loshkareva et al.
\end{figure}

\newpage
\begin{figure}
\centerline{\epsfig{file=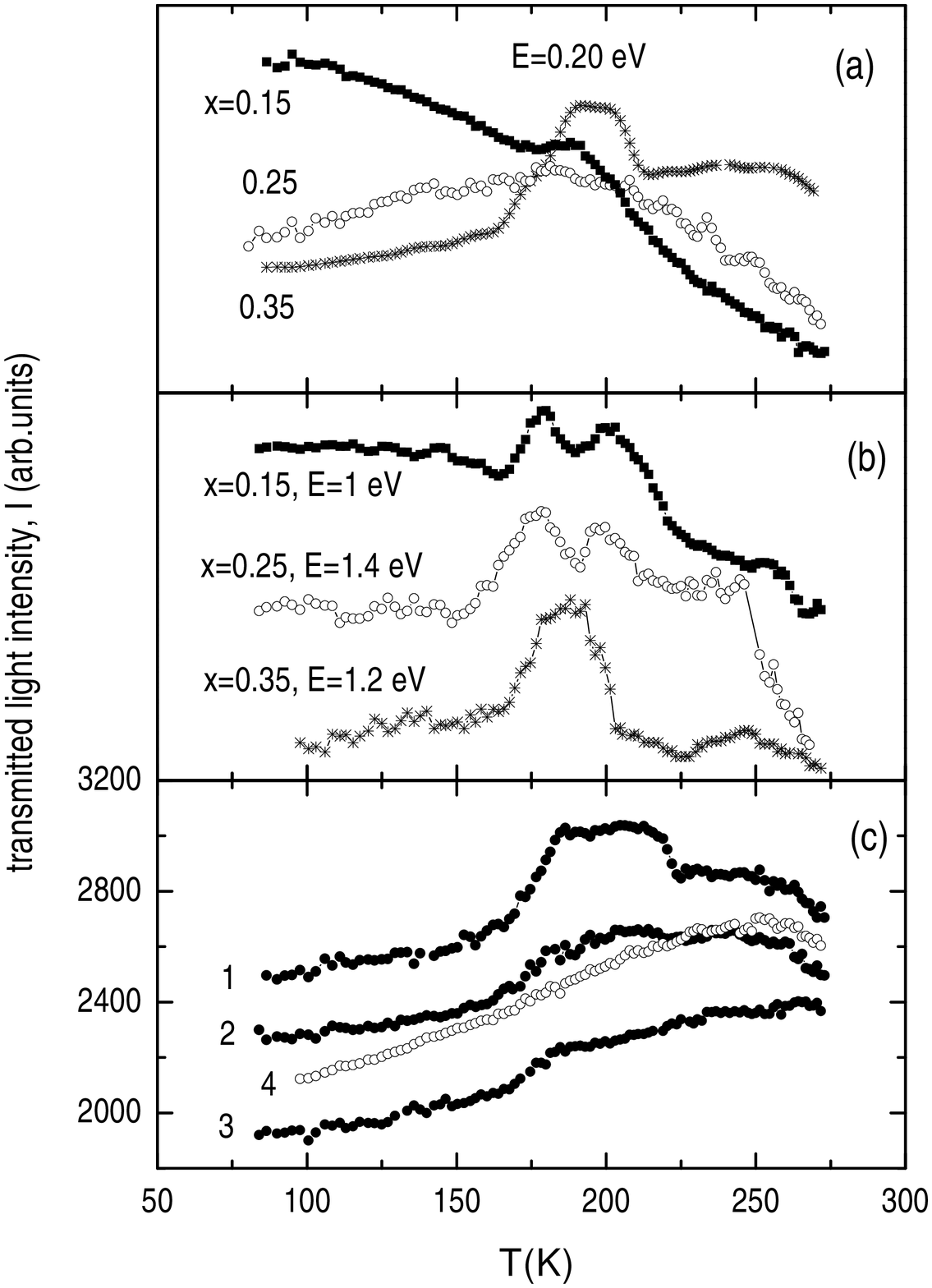,width=12cm}}
\vspace{25pt}
Figure 10 to paper Loshkareva et al.
\end{figure}

\end{document}